\documentclass[10pt,aps,prb,twocolumn,amsmath,amssymb,superscriptaddress]{revtex4-1}
\usepackage{amsmath}
\usepackage{graphicx}
\usepackage{color}
\usepackage{multirow}
\usepackage[caption = false]{subfig}
\usepackage[normalem]{ulem}

\def\beq{\begin{eqnarray}}
\def\eeq{\end{eqnarray}}
\def\beqq{\begin{eqnarray*} \color{blue} }
\def\eeqq{\end{eqnarray*}}

\def\V{\mathcal{V}}
\def\P{\mathcal{P}}

\def\Norb{{N_\mathrm{orb}}}

\begin{document}

\title{Method for near-exact energies in enormous Hilbert spaces: application to the chromium dimer}
\title{Accurate many-body electronic structure near the basis set limit: application to the chromium dimer}

\author{Junhao Li}
\affiliation{Laboratory of Atomic and Solid State Physics, Cornell University, Ithaca, NY 14853, USA}

\author{Yuan Yao}
\affiliation{Laboratory of Atomic and Solid State Physics, Cornell University, Ithaca, NY 14853, USA}

\author{Adam A. Holmes}
\affiliation{Laboratory of Atomic and Solid State Physics, Cornell University, Ithaca, NY
14853, USA}
\affiliation{Department of Chemistry and Biochemistry, University of Colorado Boulder, Boulder, CO 80302, USA}

\author{Matthew Otten}
\affiliation{Laboratory of Atomic and Solid State Physics, Cornell University, Ithaca, NY 14853, USA}

\author{Qiming Sun}
\affiliation{Tencent America LLC, Palo Alto, CA 94036, USA}
\affiliation{Division of Chemistry and Chemical Engineering, California Institute of Technology, Pasadena, CA 91125, USA}

\author{Sandeep Sharma}
\affiliation{Department of Chemistry and Biochemistry, University of Colorado Boulder, Boulder, CO 80302, USA}

\author{C. J. Umrigar}
\affiliation{Laboratory of Atomic and Solid State Physics, Cornell University, Ithaca, NY 14853, USA}

\begin{abstract}
We describe a method for computing near-exact energies for correlated
systems with large Hilbert spaces.  The method efficiently identifies the most
important basis states (Slater determinants) and performs a variational calculation in the subspace
spanned by these determinants.  A semistochastic approach
is then used to add a perturbative correction to the variational energy to compute the total energy.
The size of the variational space is progressively increased until the total energy converges
to within the desired tolerance.
We demonstrate the power of the method by computing a near-exact potential energy curve (PEC)
for a very challenging molecule -- the chromium dimer.
\end{abstract}

\maketitle

{\it Introduction:}
The evaluation of accurate energies for correlated many-electron systems is one of the most
important challenges for computational science.  The difficulty arises from the fact that the
number of many-electron states increases combinatorially with
the number of single-electron states (orbitals) $\Norb$ and the number of up- and down-spin electrons,
$N_\uparrow$, $N_\downarrow$ ($N=N_\uparrow+N_\downarrow$)
as $^\Norb C_{N_\uparrow} \times ^\Norb C_{N_\downarrow}$.

There exist a number of accurate methods for weakly correlated systems,
which we define for the purpose of this paper as systems
for which much of the wavefunction amplitude resides on a relatively small number of many-electron basis functions (Slater determinants),
all of which can be constructed by exciting electrons from the orbitals of a reference state to
orbitals within a small set of ``active" orbitals.
\footnote{We note that the usual quantum chemistry definition of weak correlation requires that
much of the amplitude resides on a single state.  Hence some systems that we consider in this paper
to be weakly correlated, would count as being strongly correlated in the quantum chemistry literature.}
In that case it is possible to perform an exact diagonalization in the complete active space (CAS),
i.e. space spanned by all determinants reachable by any number of excitations among these active orbitals.
If the orbitals are rotated to optimize the energy, the resulting method is called the
complete active space self-consistent field (CASSCF) method~\cite{WerKno-JCP-85,KreWerKno-JCP-19}.
The resulting energy can be improved by performing second-order perturbation theory to approximately include
the contribution of additional states, resulting in the complete active space perturbation theory (CASPT2) method~\cite{AndMalRooSadWol-JPC-90}.

At the other end of the spectrum, very strongly correlated systems can
be defined as those systems for which it is necessary to include
a large fraction of all possible Slater determinants
to get an accurate energy and other expectation values.
For these systems there is no recourse other than exact diagonalization in the entire Hilbert space,
which is feasible only for very small systems or very small basis sets.

In between these two extremes, moderately strongly correlated systems require
a large number of important Slater determinants
to obtain accurate results, but this number constitutes a vanishingly small fraction
of the dimension of the Hilbert space.
Further, these states do not have any obvious pattern (e.g. they do not all belong to a CAS space).
Many {\it ab initio} Hamiltonians belong to this category.
It is for these systems that selected configuration interaction plus perturbation theory (SCI+PT),
first developed about 50 years ago~\cite{BenDav-PR-69,HurMalRan-JCP-73}, can be most useful.
Recently there has been renewed interest in these
methods~\cite{Eva-JCP-14,LiuHof-JCTC-16,SceAppGinCaf-JCoC-16,TubLeeTakHeaWha-JCP-16,GarSceLooCaf-JCP-17,DasMorSceFil-JCTC-18,GarSceGinCafLoo-JCP-18,LooSceBloGarCafJac-JCTC-18,HaiTubLevWhaHea-JCTC-19}
and some interesting applications, particularly to excited states~\cite{LooSceBloGarCafJac-JCTC-18,ChiHolOttUmrShaZim-JPCA-18}.
The recent development of a very efficient algorithm in the form of the semistochastic heatbath configuration interaction (SHCI)
method by some of the authors of this paper~\cite{HolTubUmr-JCTC-16,ShaHolJeaAlaUmr-JCTC-17,HolUmrSha-JCP-17,SmiMusHolSha-JCTC-17,LiOttHolShaUmr-JCP-18} has now made it possible
to perform calculations on a wider and more interesting set of systems.
We next briefly describe the SHCI method and the main innovations that account for its efficiency.
Then we apply the SHCI method to calculate the potential energy curve of a
small but very challenging molecular system, the chromium dimer.

{\it Method:}
Selected configuration interaction plus perturbation theory (SCI+PT) methods
approximate the full configuration interaction (FCI) energy by selecting
the most important determinants from a large Hilbert space.
These methods contain two steps.
In the first step a set of important determinants, $\V$, are selected and the Hamiltonian is diagonalized in the subspace of these
determinants to obtain the
lowest, or the lowest few, eigenstates.
In the second step, a second-order perturbation theory is used to calculate the energy contributions
of the
determinants that are in a space P that is disjoint to V, but that
have a non-zero Hamiltonian matrix element connecting them to at least one of the determinants in $\V$.
We will refer to $\V$ and $\P$ as the variational and perturbative spaces, respectively.
The recently developed SHCI algorithm substantially reduces the computational time of performing both the variational calculation and the
perturbative correction, and eliminates the memory bottleneck for the perturbative calculation.
We describe these two innovations next.

The method used in this paper is an improved version of the one recently developed~\cite{LiOttHolShaUmr-JCP-18} by some of the authors of this paper.
Straightforward SCI+PT implementations
use an energetic criterion based on $2^{nd}$-order perturbation theory,
\beq
\frac{\left(\sum_{D_i \in \mathcal{V}} H_{ai} c_i\right)^2}{E - E_a} < -\epsilon,
\label{eq:cipsi}
\eeq
for selecting determinants, $D_a$, to be included in $\V$.
In Eq.~\ref{eq:cipsi}, $E$ is the energy of the variational wavefunction and $E_a$ is the energy of determinant $D_a$.
SHCI modifies the selection criterion to
\beq
\max_{D_i \in \mathcal{V}}\left|  H_{ai} c_i \right| > \epsilon_1,
\label{eq:shci}
\eeq
which greatly reduces the cost by taking
advantage of the fact that most of the $H_{ai}$ matrix elements are 2-body excitations, which
depend only on the indices of the 4 orbitals whose occupations change and not on the other occupied orbitals of a determinant~\cite{HolTubUmr-JCTC-16}.
Thus by sorting the absolute values of all possible matrix elements of the 2-body excitations in descending order prior to the start of the SHCI selection algorithm,
the scan over determinants $D_a$ can be terminated as soon as $|H_{ai}|$ drops below $\epsilon_1/|c_i|$.
In this paper, a similar idea is used to speed up the selection of 1-body excitations as well.
This enables a procedure in which \emph{only the important determinants are ever looked at}, resulting in
orders of magnitude saving in computer time.
SHCI uses an analogous procedure to also efficiently select the important perturbative determinants in $\P$, replacing the variational cutoff $\epsilon_1$ in Eq. (2) with a much smaller perturbative cutoff $\epsilon_2 \ll \epsilon_1$.

Even with this improvement, a straightforward evaluation of the perturbative correction has a very large memory requirement because all distinct determinants
in $\P$
that meet the criterion
in Eq.~\ref{eq:shci} with $\epsilon_1$ replaced by $\epsilon_2$
must be stored.
\footnote{An alternative straightforward approach does not have a large memory requirement, but
requires considerably larger computation time.}
For the systems in this paper, the total number of connected determinants is $> 10^{15}$ ($> 10^{13}$ \emph{distinct} connected determinants)
when the number of variational determinants is on the order of $10^9$.
To solve this problem, we have developed a 2-step~\cite{ShaHolJeaAlaUmr-JCTC-17}, and later an improved
3-step~\cite{LiOttHolShaUmr-JCP-18} semistochastic perturbative approach
that overcomes this memory bottleneck, and is fast and perfectly parallelizable.
A different efficient semistochastic perturbative approach has been used in Ref.~\onlinecite{GarSceLooCaf-JCP-17}.
These improvements allowed us to use $2 \times 10^9$ variational determinants~\cite{LiOttHolShaUmr-JCP-18},
which is two orders of magnitude larger than the largest variational space of $2 \times 10^7$
determinants~\cite{GarSceLooCaf-JCP-17} used in any other SCI+PT method.

We choose $\epsilon_2 = 10^{-6} \epsilon_1$, so by progressively reducing the single parameter
$\epsilon_1$ a systematic convergence to the full configuration interaction limit is obtained.
The energy at the $\epsilon_1=0$ limit is obtained using a quadratic fit to the energies
versus the perturbative correction~\cite{HolUmrSha-JCP-17}.
The convergence of the energy depends greatly on the choice of orbitals.
Natural orbitals give faster convergence than Hartree Fock orbitals.
Orbitals that are optimized to minimize the SHCI energy~\cite{SmiMusHolSha-JCTC-17} for a large value of $\epsilon_1$
yield yet faster convergence, but the optimization typically requires many more optimization iterations than CASSCF optimizations require because of strong coupling between the orbital and CI parameters.
In this paper we greatly accelerate the convergence by
using an overshooting method based on the angle between successive parameter updates.

{\it Potential energy curve of Cr$_2$:}
The potential energy curve of the chromium dimer is very challenging for state-of-the-art quantum chemistry methods for several reasons.
The $^1\Sigma_g^+$ ground state of the molecule dissociates into two atoms in high-spin $^7S$ states, each with 6 unpaired 3d and 4s electrons.
Thus the molecule has a formal sextuple bond, and the minimal CAS space required for correct dissociation is CAS(12e,12o).
Consequently, near-degeneracy correlation is very important, as evidenced by the fact that spin-unrestricted coupled cluster theory
with single, double and perturbative triple excitations (UCCSD(T)) predicts a dissociation energy
that is much too small~\cite{Bauschlicher1994}.
Simultaneously, dynamic correlation is also very important, as evidenced by the fact that CASSCF in a CAS(12e,12o) space gives a very weak
minimum at a very large bond length.
Thus, most of the calculations that have been performed employ CASPT2~\cite{KurYan-JCP-11,RuiAquUgaInf-JCTC-11,MaManOlsGag-JCTC-16,VanMalVer-JCTC-16} or the related
n-electron valence state perturbation theory (NEVPT2)~\cite{Guo2016} to try to capture both near-degeneracy and dynamic correlation effects.
These methods are sensitive to the choice of the CAS space, and in addition the CASPT2 method is sensitive
to the choice of the ionization potential electron affinity (IPEA) shift.
In fact, CASPT2 with a CAS(12e,12o) reference space and reasonable choices of IPEA shift yield well depths ranging from 1.1 to 2.4 eV~\cite{MaManOlsGag-JCTC-16}.
Since conventional CASSCF calculations are limited to about CAS(18e,18o), the density matrix renormalization group
(DMRG)~\cite{Whi-PRL-92,ChaDorGhoHacNeuWanYan-FQSCP-09} method has been employed~\cite{KurYan-JCP-11,Guo2016} as a
CAS space solver, allowing the use of the larger CAS(12e,22o), CAS(12e,28o), and CAS(28e,20o) reference spaces, which
partially cures this problem.
Despite this, these methods have been unable to provide a definitive potential energy curve (PEC) for Cr$_2$.

Externally contracted multireference configuration interaction (MRCI) using determinants
selected from a DMRG calculation in a CAS(12e,42o) as the reference space has also been used~\cite{LuoMaWanMa-JCTC-18}.
It gives a reasonably-shaped PEC shifted down by about 0.1 eV relative to experiment.
Multireference averaged quadratic coupled cluster (MR-AQCC)~\cite{Mul-JPCA-09} is another accurate method that
has been used to compute the PEC of Cr$_2$.  It gives a well depth of 1.35 eV, and the shape of the PEC is in
reasonable agreement with experiment.

Probably the most accurate method used for Cr$_2$ is the Auxiliary Field Quantum Monte Carlo (AFQMC)~\cite{PurZhaKra-JCP-15}.
Most AFQMC computations are performed using the phaseless approximation, the accuracy of which depends on the choice
of the trial wavefunction.  For Cr$_2$, phaseless AFQMC is not sufficiently accurate with affordable trial wavefunctions.
On the other hand free-projection AFQMC is exact (aside from statistical error), but very computationally expensive.
So, a hybrid approach was used wherein free-projection AFQMC was performed in the 3z basis and
the complete basis set correction was computed by adding in the correction from phaseless AFQMC for 3z, 4z, and 5z
basis sets for $r < 2$ \AA, and by adding in the correction from
free-projection AFQMC with only 12 rather than 28 correlated electrons in 3z and 4z basis sets for $r > 2$ \AA.

Part of the interest in Cr$_2$ comes from the fact that an experimentally deduced PEC is available which can be used
to some extent to test the accuracy of theoretical methods.
The shape of the PEC comes from high-resolution photoelectron spectra of Cr$_2^-$, which showed 29 vibrationally
resolved transitions to the neutral Cr$_2$ ground state~\cite{CasLeo-JPC-93}.  However, there are gaps in the measured vibrational
levels and the assignment of the higher levels is not unambiguous, so part of the PEC is not well constrained
by the data.
The vertical placement of the potential energy curve is determined from
the dissociation energy.  Experimental values vary considerably:
1.56(26)~\cite{KanStr-JCP-66}, 1.78(35)~\cite{KanStr-JCP-66}, 1.44(6)~\cite{HilRut-BBPC-87},
1.43(10)~\cite{SuHalArm-CPL-93}, and 1.54(6)~\cite{simard1998photoionization} eV.
We will use the last number in most of our plots, since it is more recent, but will keep in mind that it
has considerable uncertainty.
Since the zero point energy is 0.03 eV~\cite{KurYan-JCP-11}, the potential energy curves we present are shifted
so that the well depth is 1.57 eV.
Recently, the experimental data of Casey and Leopold~\cite{CasLeo-JPC-93} have reanalysed by Dattani~\cite{Dat-unpub-17} using a more flexible fitting function
and a fully quantum mechanical treatment to obtain a slightly different PEC from the original.
We show both of these curves in all our figures.

{\it Hamiltonian:}
For the 3d transition metals it is important to include scalar relativistic effects, but the spin-orbit splitting is small.
The two standard scalar relativistic Hamiltonians are the Douglas-Kroll and the {\it x2c}~\cite{KutLiu-JCP-05} Hamiltonians.
In our work we employ mostly the {\it x2c} Hamiltonian, but we have verified that the Douglas-Kroll Hamiltonian yields
essentially the same PEC, though it gives a total energy for the molecule that is about 14.6 mHa higher.
The 1- and 2-body integrals for the {\it x2c} Hamiltonian are obtained using the PySCF package~\cite{SunCha_etal_PySCF-ComMolSci-18}.

{\it Basis sets:}
Quantum chemists have designed several different sets of standard single-particle basis functions
for most of the elements in the periodic table~\cite{BasisSetExchange}.
The ``correlation consistent" bases of Dunning and coworkers~\cite{Dun-JCP-89,BalPet-JCP-06,SchDidElsSunGurChaLiWin-JCIM-07}
are widely used and are designed to enable systematic extrapolation to the complete basis limit.
These bases are designated cc-pV$n$Z, where $n$ is referred to as the cardinal number of the basis set.
They are designed for non-relativistic calculations; the corresponding basis sets for
relativistic calculations are designated cc-pV$n$Z-DK.
We employ the cc-pV$n$Z-DK basis sets with $n$ ranging from 2-5, and for brevity we designate these by 2z, 3z, 4z and 5z.
These have 86, 136, 208 and 306 basis functions for the dimer, respectively, which result in the same number of orbitals
written as linear combinations of the basis functions.
In the SHCI calculations we allow excitations to and from all these orbitals, keeping only a small number of core,
and in some calculations semicore, orbitals doubly occupied.
By using more than one basis set, we can extrapolate the UCCSD(T) and SHCI energies
to the complete basis limit making the usual assumption that the
binding energy converges as the inverse cube of the cardinal number, $n$, for $n \ge 3$.

Cr$_2$ at a bond length of $1.5$ \AA\ in the Ahlrichs SV basis~\cite{SchHorAhl-JCP-92}
has become a very popular system for
testing the accuracy and efficiency of electronic structure methods, even though this
basis is much too small to give even a qualitatively correct potential energy curve~\cite{KurYan-JCP-09}.
In the Supplementary Material~\cite{Supplementary_Cr2}, we provide accurate energies for this basis, both with and
without core excitations.

{\it Correlating 12 electrons:}
Molecular systems containing heavy atoms have orbitals with very different energies.
Although core electron correlations make a large contribution to the total energy,
they have only a relatively small effect on
energy differences such as the potential energy curve (PEC) because the core contributions in the atoms
and the molecule tend to cancel.  In Cr, the 3d and 4s electrons are the valence electrons,
the 3s and 3p electrons are semicore electrons, and the 1s, 2s and 2p electrons are the core electrons.
Early calculations of Cr$_2$ employed only valence electron excitations; later calculations included also
semicore electron excitations.

\begin{figure}[htb]
  \begin{center}
  \includegraphics[width=0.9\linewidth]{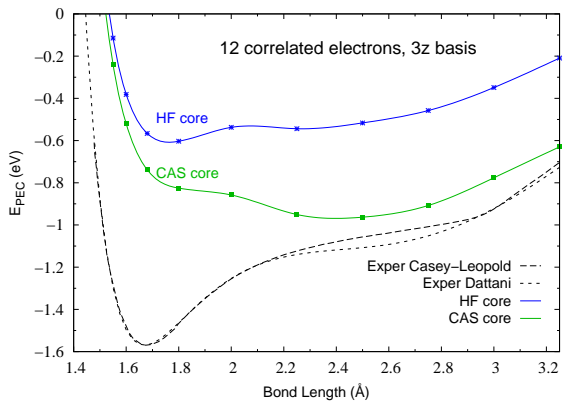}
  \caption{Comparison of the SHCI potential energy curves correlating the 12 valence electrons
  with a HF core and a CAS core to experimentally deduced curves.
  %SHCI allows excitations to all higher-lying orbitals.
  Note that in the SHCI calculation excitations to all higher-lying orbitals are allowed.
  When correlating 12 electrons the nature of the frozen orbitals has a large effect on the PEC.
  }
  \label{fig:12correlated_3z}
  \end{center}
\end{figure}

The computed energies depend not only on which orbitals are allowed to excite, but also on the nature
of the orbitals that are kept frozen (not allowed to excite).
Fig.~\ref{fig:12correlated_3z} shows the PEC obtained from correlating only the 12 valence electrons
by allowing excitations to all higher lying orbitals,
keeping the semicore and core electrons fixed either in Hartree-Fock (HF) orbitals, or
in orbitals obtained by optimizing in a CAS(12e,12o) space.
The two curves differ greatly from each other and from the experimentally deduced PECs.

\begin{figure}[htb]
  \begin{center}
  \includegraphics[width=9cm]{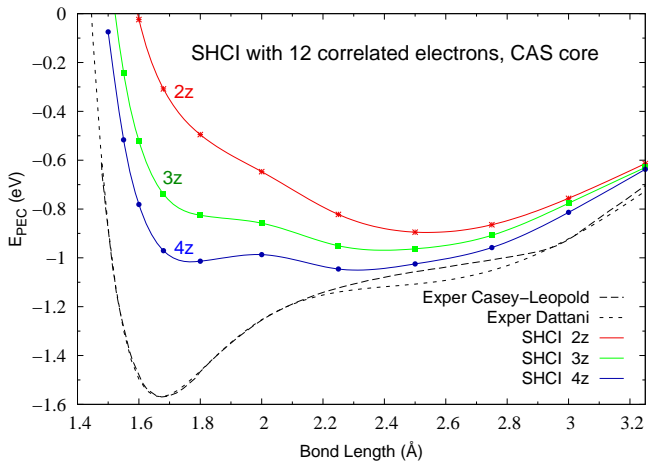}
  \caption{Comparison of the SHCI potential energy curves, correlating the 12 valence electrons
  using a CAS core and 2z-4z basis sets, to experimentally deduced curves.  It is apparent that correlating
  just the 12 valence electrons is insufficient to get an accurate PEC.
  }
  \label{fig:12correlated_CAS_core}
  \end{center}
\end{figure}

In Fig~\ref{fig:12correlated_CAS_core} we employ the 2z, 3z and 4z basis sets to study the basis set dependence of the PECs obtained again from
correlating only the 12 electrons, using CAS(12e,12o) semicore and core orbitals.
Although the PECs improve with increasing basis size, it is clear that
correlating just 12 electrons is insufficient to get good agreement with experiment.
This is in fact well known, but the precise PECs have not been published before.

\begin{figure}[htb]
  \begin{center}
  \includegraphics[width=9.1cm]{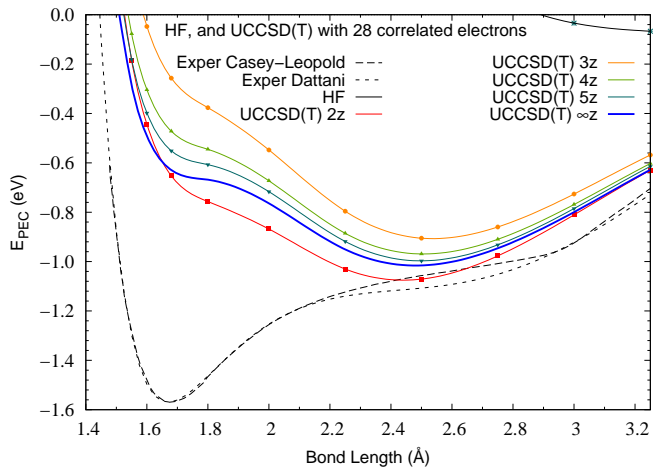}
  \caption{UHF and UCCSD(T) potential energy curves correlating 28 electrons in bases ranging from
  2z to 5z, and the complete basis limit.
  %curves correlating 28 and 44 electrons.
  Note that the 2z curve lies lower than the 3z, 4z over the entire range, and lower than
  the 5z and complete basis curves over most of the range.
  }
  \label{fig:UCCSDT}
  \end{center}
\end{figure}

\begin{figure}[htb]
  \begin{center}
  \includegraphics[width=9cm]{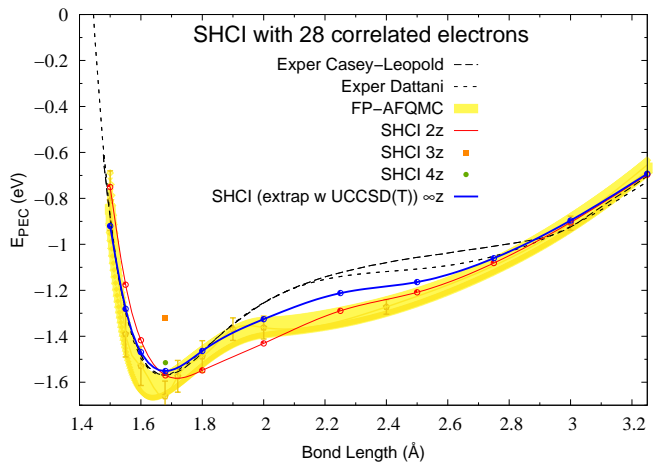}
  \caption{Comparison of the SHCI potential energy curves correlating 28 electrons
  to experimentally deduced curves.  The red curve is for the 2z basis, the orange dot for the
  3z basis, the green dot for the 4z basis and the blue curve is % the extrapolation to
  the complete basis limit using the correction from UCCSD(T).
  Similarly to UCCSD(T), when 28 electrons are correlated, the binding energies
  do not change monotonically with the basis cardinal number.
  The FP-AFQMC curve from Fig. 4 of Ref.~\onlinecite{PurZhaKra-JCP-15} is also shown.
  %The dashed blue curve is for 44 correlated electrons (it includes 2s and 2p core correlations)
  %extrapolated to the complete basis limit.
  }
  \label{fig:28correlated_CAS_core}
  \end{center}
\end{figure}

{\it Correlating 28 electrons:}
The coupled cluster method with single, double and perturbative triples (CCSD(T)) amplitudes gives very
accurate energies for
systems where a single determinant has a large amplitude,
such as most organic molecules at equilibrium geometry.
Here we use the spin-unrestricted versions of HF and CCSD(T), denoted by UHF and UCCSD(T) respectively
meaning that the HF up-spin and down-spin orbitals,  and the CCSD up-spin and down-spin amplitudes,
need not be the same, since this allows for dissociation of the molecule into two high-spin atoms.
On the other hand, in our SHCI calculations, up- and down-spin orbitals are the same,
so that the SHCI wavefunction can be an eigenstate of $S^2$.
In Fig.~\ref{fig:UCCSDT} we show the PECs obtained from UCCSD(T)
using PySCF~\cite{SunCha_etal_PySCF-ComMolSci-18} and 2z through 5z basis functions.
Of course, the total energies go down monotonically with increasing basis size, but
very surprisingly the 2z PEC curve lies lower than the 3z, 4z and 5z curves.
The same behavior is observed also in SHCI calculations at equilibrium with 2z, 3z, and 4z bases.
The infinite basis extrapolated UCCSD(T) curve, shown as the solid blue line lies below the 2z
curve at short distances and above the 2z curve at large distances.
The extrapolation is done using the 4z and 5z curves, but almost the same extrapolated curve is
obtained from 3z and 4z curves.
The 28 correlated electron UCCSD(T) curves have shapes similar to those from the 12 correlated
electron SHCI curves, but they agree even less well with experiment.

Although UCCSD(T) gives poor PECs, it can be used to provide a reasonable basis set correction
to the SHCI curves that we present next.  The accuracy of the correction has been checked
at the equilibrium bond length, where we find that the
correction to the 2z SHCI energy, using the (3z,4z,5z) UCCSD(T) energies agrees with that obtained from
the (3z,4z) SHCI energies within 0.1 eV.
More precise agreement cannot be expected since the uncertainty of the SHCI 3z, 4z energies from the extrapolation in $\epsilon_1$
is itself about 0.1 eV.
The 4z SHCI calculations with 28 correlated electrons have a Hilbert space of $(^{198}C_{14})^2 \approx 10^{42}$.
One of the desirable features of the SHCI method is that although the Hilbert space increases
by 10 orders of magnitude going from the 2z to the 4z basis, the cost of the calculation is only
a few times larger.  This desirable feature is even more evident when the increase in Hilbert space size comes from
correlating additional core orbitals.  However, since the 2z calculations are already expensive, we have
done the larger basis calculations only at equilibrium.

The PEC from SHCI in the 2z basis, correlating the 28 valence and semicore electrons is shown as the red curve
in Fig.~\ref{fig:28correlated_CAS_core}.  The blue curve is the PEC extrapolated to infinite
basis size using the correction from UCCSD(T).
It has a minimum of -1.55 eV at 1.679 \AA, in agreement with the experimentally determined
-1.57 eV~\cite{simard1998photoionization} at 1.679 \AA~\cite{CasLeo-JPC-93}.
It agrees very well with experiment at bond lengths around equilibrium and also at long bond lengths.
It differs a little from experiment in the shoulder region from 1.8 to 2.7 \AA, which roughly coincides
with the range of distances where the experimentally deduced curve is most uncertain because
of missing vibrational levels, as also noted in Ref.~\onlinecite{PurZhaKra-JCP-15}.
This is also the region where the computed energies converge most slowly.
The blue curve agrees well also with the curve labeled FP-AFQMC in Fig. 4 of Ref.~\onlinecite{PurZhaKra-JCP-15},
except that the FP-AFQMC curve is yet a bit lower than SHCI in the shoulder region.

{\it Conclusions:}
The SHCI method enables systematic convergence to the exact energy for moderately strongly
correlated systems with sizes of Hilbert space that were previously inaccessible.
We have demonstrated its power by computing the potential energy curve of a very challenging molecule, Cr$_2$.
The size of the largest Hilbert space treated with SHCI is $10^{42}$ determinants.
Nevertheless,
energies estimated to be accurate to a few milliHartrees
were obtained from calculations that involve $10^9$ variational determinants or fewer,
and several trillion perturbative determinants.
In future work we plan to use an effective Hamiltonian that incorporates the effect of explicit interelectronic
correlation~\cite{YanShi-JCP-12} to reduce the magnitude of the basis set extrapolation error.

\begin{acknowledgements}
This work was supported by the AFOSR under grant FA9550-18-1-0095 and by the NSF under grants ACI-1534965
and CHE-1800584.
The computations were performed on the Bridges cluster at the Pittsburgh Supercomputing
Center supported by NSF grant ACI-1445606, as part of the XSEDE program supported by NSF grant
ACI-1548562, and on the Google Cloud Platform.
We thank Nike Dattani for sharing the Cr$_2$ PEC he deduced from experimental data,
Ankit Mahajan for help with using PySCF,
and Garnet Chan, Andreas Savin and Julien Toulouse for valuable discussions.
\end{acknowledgements}

\bibliographystyle{jchemphys}
%\bibliography{all}

\end{document}

\title{Supplementary material for: \\ Accurate many-body electronic structure near the basis set limit: application to the chromium dimer}

\author{Junhao Li}
\affiliation{Laboratory of Atomic and Solid State Physics, Cornell University, Ithaca, NY 14853, USA}

\author{Yuan Yao}
\affiliation{Laboratory of Atomic and Solid State Physics, Cornell University, Ithaca, NY 14853, USA}

\author{Adam A. Holmes}
\affiliation{Laboratory of Atomic and Solid State Physics, Cornell University, Ithaca, NY
14853, USA}
\affiliation{Department of Chemistry and Biochemistry, University of Colorado Boulder, Boulder, CO 80302, USA}

\author{Matthew Otten}
\affiliation{Laboratory of Atomic and Solid State Physics, Cornell University, Ithaca, NY 14853, USA}

\author{Qiming Sun}
\affiliation{Tencent America LLC, Palo Alto, CA 94036, USA}
\affiliation{Division of Chemistry and Chemical Engineering, California Institute of Technology, Pasadena, CA 91125, USA}

\author{Sandeep Sharma}
\affiliation{Department of Chemistry and Biochemistry, University of Colorado Boulder, Boulder, CO 80302, USA}

\author{C. J. Umrigar}
\affiliation{Laboratory of Atomic and Solid State Physics, Cornell University, Ithaca, NY 14853, USA}

\maketitle

\section{Energy convergence in Ahlrichs SV basis} \vspace{-3mm}
For Cr$_2$, the Ahlrichs SV basis~\cite{SchHorAhl-JCP-92} (somewhat confusingly named Ahlrichs VDZ basis at the Basis Set Exchange~\cite{BasisSetExchange})
is too small to give even a qualitatively correct potential energy curve~\cite{KurYan-JCP-09}.
The predicted equilibrium bond length is much too long.
However, Cr$_2$ at a bond length of $1.5$ \AA\ in this basis
has become a very popular system for
testing the accuracy and efficiency of electronic structure
methods~\cite{KurYan-JCP-09,ShaCha-JCP-12,BooSmaAla-MP-14,EriGau-JCPL-19,GuoLiCha-JCP-18,GuoLiCha-JCTC-18,OliHuNakShaYanCha-JCP-15,TubLeeTakHeaWha-JCP-16,HolTubUmr-JCTC-16,XuUejTen-PRL-18,TubFreLevHaiHeaWha-ARX-18,ZhaLiuHof-ARX-19}.
Since the SHCI method has progressed considerably after publishing our earlier
calculations for this basis, we provide updated information here.

\begin{figure}[!h]
  \begin{center}
  \includegraphics[width=0.77\linewidth]{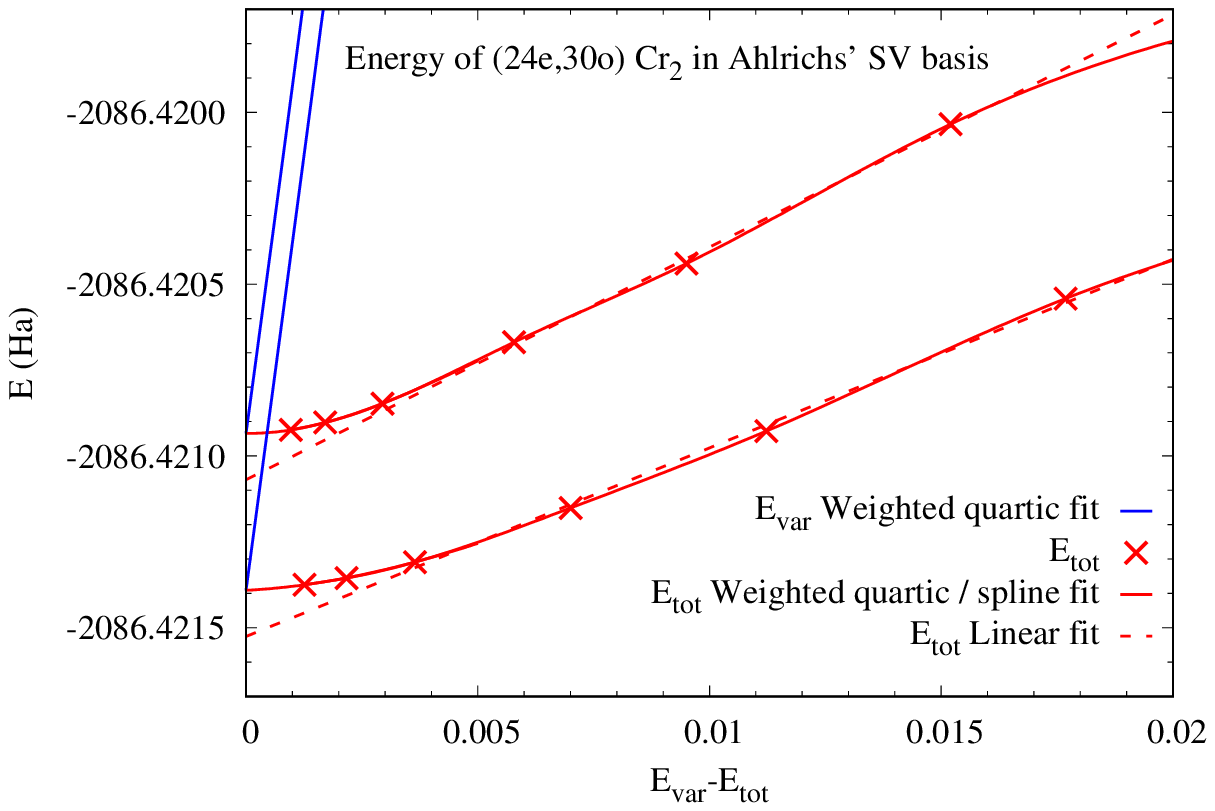}
  \caption{Convergence of the frozen Mg-core (24e,30o) total and variational energies.
  The upper pair of curves freeze the core in HF orbitals and the lower pair in CAS(12,12) natural orbitals.
  }
  \label{fig:24e_Ahlrichs}
  \end{center}
\end{figure}

\begin{figure}[!h]
  \begin{center}
  \includegraphics[width=0.77\linewidth]{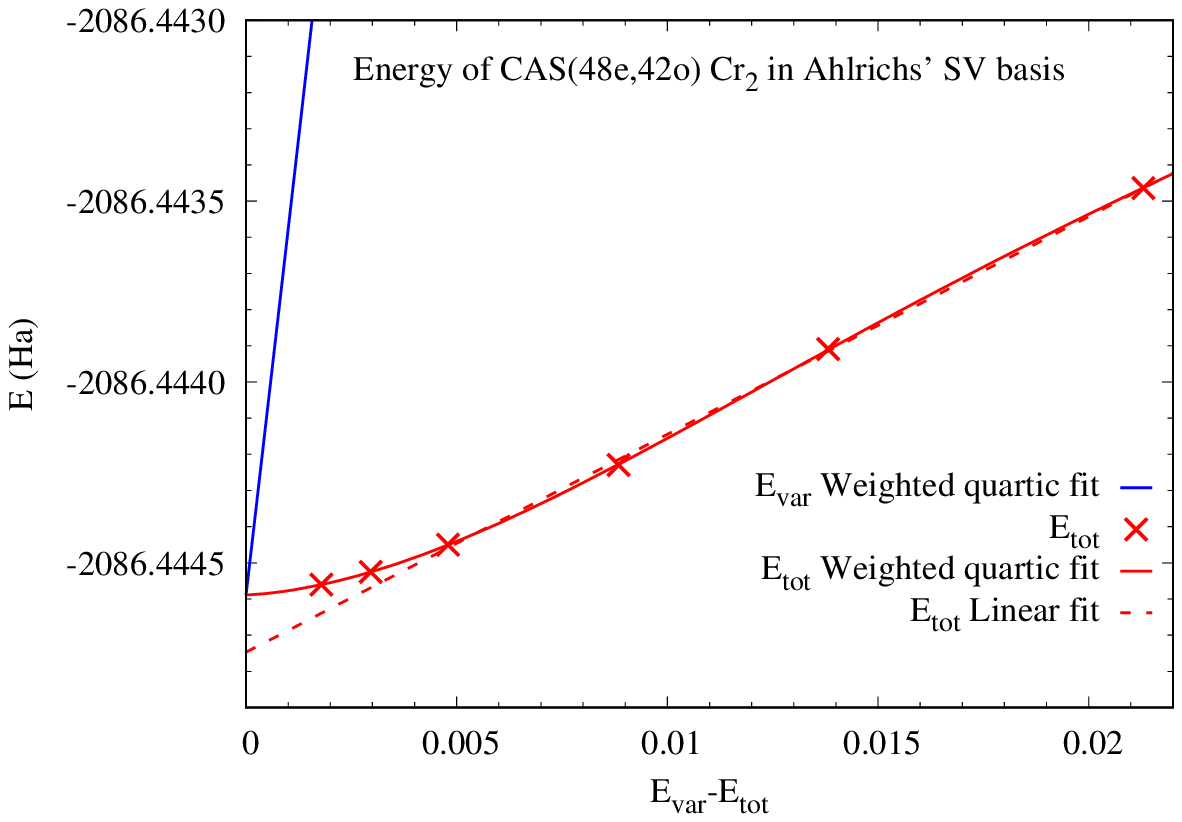}
  \caption{Convergence of the all-electron (48e,42o) total and variational energies.
  }
  \label{fig:48e_Ahlrichs}
  \end{center}
\end{figure}

Table~\ref{tab:Ahlrichs} and Figs.~\ref{fig:24e_Ahlrichs} and \ref{fig:48e_Ahlrichs}
show the convergence of the energy, for both frozen Mg-core calculations (excitation space of $(24e,30o)$), and
for all electron excitation calculations (excitation space of $(48e,42o)$).
The (24e,30o) energies depend on the nature of the frozen core orbitals.
We studied freezing Hartree-Fock (HF) core orbitals and freezing core natural orbitals from
a CAS(12e,12o) space.
To get the energy extrapolated to $\epsilon_1=0$, rather than fitting to a polynomial in $\epsilon_1$,
it is preferable~\cite{HolUmrSha-JCP-17} to fit to a polynomial in $E_{\rm var}-E_{\rm tot}$.
In Figs.~\ref{fig:24e_Ahlrichs} and \ref{fig:48e_Ahlrichs} the solid lines are weighted quartic fits,
using $1/(E_{\rm var}-E_{\rm tot})^2$ as the weight function.  In some curves a spline fit is used
for part of the range, but these coincide with the quartic fits within the thickness of the lines over almost all the range.
We also show linear fits (dashed lines) using just the 4 points with $\epsilon_1$ ranging from $2 \times 10^{-5}$ to $2 \times 10^{-4}$ Ha
(the largest 4 values of $\epsilon_1$ shown in the plots) merely to demonstrate the error resulting
from not going to sufficiently small values of $\epsilon_1$ (sufficiently large $N_{\rm det}$).
From the figures, it is apparent that the weighted quartic fits provide more accurate extrapolated energies.
The quartic fits are possible only because we went down to sufficiently small values of $\epsilon_1$ and because the statistical
uncertainties in our calculations are very small, particularly for the smaller $\epsilon_1$ values.
The Table shows that the iCIPT2 energies in a very recent preprint~\cite{ZhaLiuHof-ARX-19} agree very well
with our linear extrapolations, but not as well with the more accurate weighted quartic extrapolation.
The extrapolated energies in Table~\ref{tab:Ahlrichs} for HF-core $(24e,30o)$, CAS-core $(24e,30o)$ and $(48e,42o)$ should be
accurate to 0.005, 0.005 and 0.01 mHa respectively and should serve as a reference for other methods.
In case the reader is surprised that the weighted quartic fits for $E_{\rm var}$ and $E_{\rm tot}$ extrapolate
to precisely the same point, we note that the expansion coefficients for the two fits are precisely
the same, except that the linear coefficients differ by 1.

\begin{figure}[!h]
  \begin{center}
  \includegraphics[width=0.8\linewidth]{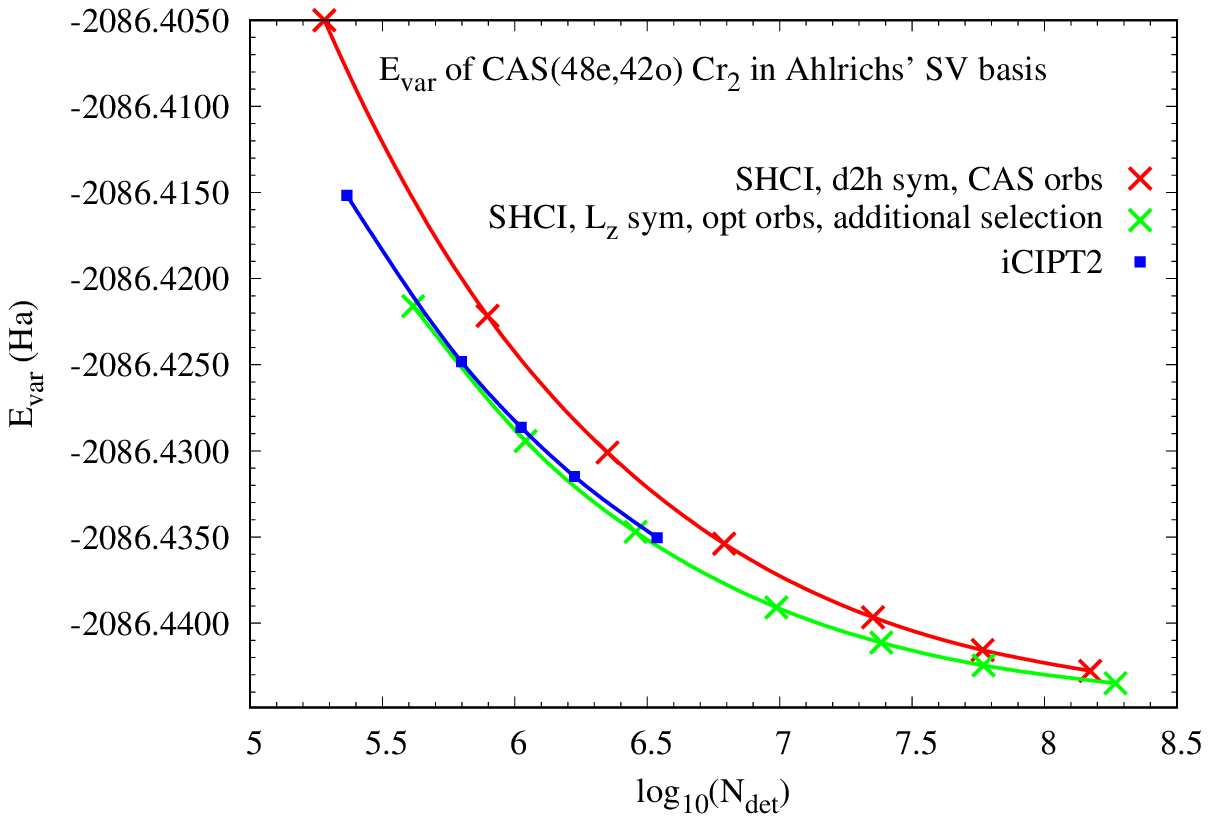}
  \caption{Demonstration of improvement in convergence of the (48e,42o) variational energy
  upon using $L_z$ symmetry, optimized orbitals and the additional selection criterion.
  The iCIPT2 energies of Ref.~\onlinecite{ZhaLiuHof-ARX-19} are also shown.
  }
  \label{fig:newselection}
  \end{center}
\end{figure}

The convergence of the energies versus the number of determinants can be improved by using orbitals
with $L_z$ rather than d2h point group symmetry, and by optimizing the orbitals~\cite{SmiMusHolSha-JCTC-17}.
This is the reason why in Table~\ref{tab:Ahlrichs}, for a given value of $N_{\rm det}$,
the energies in the first block are better converged than those in the second block,
each relative to its converged value.
Besides these two changes, there is yet another improvement
that can be made to the convergence of the (48e,42o) calculations.  The usual SHCI selection criterion in Eq. 2 of the main paper
does not take into account the large differences in the energy denominator of
$2^{nd}$-order perturbation theory when core excitations are allowed.  An efficient way to remove unimportant high excitation energy determinants
is to add a second selection criterion,
\beq
\frac{\left(H_{ai} c_i\right)^2}{\max\left(\sum_i e_{a,i}-\sum_i e_{{\rm HF},i}, 0 \right)} > c\epsilon^2,
\label{eq:new_selection}
\eeq
where $e_{{\rm HF},i}$ and $e_{a,i}$ are the 1-body energies of the $i^{th}$ occupied orbital in the HF determinant and in determinant
$D_a$ respectively.
If $c=0$, the additional selection criterion has no effect.  We used $c=0.2$.
The resulting improvement in the convergence is shown in Fig.~\ref{fig:newselection}.
For this system, SHCI with L$_{\rm z}$ optimized orbitals and the additional selection criterion converges slightly faster
than iCIPT2~\cite{ZhaLiuHof-ARX-19}, but more importantly SHCI can go to much larger $N_{\rm det}$.

\newpage

\setlength\tabcolsep{3pt}
\begin{table}[!h]
\begin{threeparttable}
\caption{Convergence of Cr$_2$ variational and total energies at r=1.5 \AA, in the (24e,30o) and (48e,42o) spaces using the Ahlrichs SV basis
versus $\epsilon_1$ (see Eq. 2 and surrounding text of the main paper).
The energy in the (24e,30o) space is 0.46 mHa deeper if the frozen core orbitals are CAS(12,12) natural orbitals rather than Hartree-Fock (HF) orbitals.
Most papers in the literature appear to use the CAS(12,12) core, but some papers do not specify.
$N_{\rm det}$ is the number of determinants in the variational wavefunction.
The estimated statistical and/or extrapolation errors in the last digit(s) of $E_{\rm tot}$ are in parentheses.
The most accurate extrapolation is shown in bold face.
In both spaces, the iCIPT2~\cite{ZhaLiuHof-ARX-19} energies agree very well with the linear extrapolations,
but less well with the more accurate weighted quadratic extrapolations.
The timings shown are for 2 Intel Xeon E5-2620 v4 nodes, each with 16 physical cores running at 2.1 Gz.}
\label{tab:Ahlrichs}
\begin{tabular}{ l r l l d d }
\hline \hline
& & \multicolumn{2}{c}{Energy + 2086 (Ha)} & \multicolumn{2}{c}{Real time (sec)} \\
$\epsilon_1$ (Ha) & $N_{\rm det}$ & \multicolumn{1}{c}{$E_{\rm var}$} & \multicolumn{1}{c}{$E_{\rm tot}$} & \multicolumn{1}{c}{var} & \multicolumn{1}{c}{PT2} \\
\hline
& & \multicolumn{2}{c}{Excit. space (24e,30o)} \\
& & \multicolumn{2}{c}{Hartree-Fock core     } \\
& & \multicolumn{2}{c}{Optimized L$_{\rm z}$ orbitals } \\
$5 \times 10^{-4}$ & 113\,322      &  -0.390\,528    & -0.419\,564(10)  & 1 & 37 \\
$2 \times 10^{-4}$ & 429\,970      &  -0.404\,833    & -0.420\,035(8)   & 5 & 53 \\
$1 \times 10^{-4}$ & 1\,108\,805   &  -0.410\,938    & -0.420\,440(6)   & 11 & 76 \\
$5 \times 10^{-5}$ & 2\,854\,759   &  -0.414\,887    & -0.420\,669(4)   & 36 & 128 \\
$2 \times 10^{-5}$ & 9\,505\,470   &  -0.417\,904    & -0.420\,848(1)   & 183 & 416 \\
$1 \times 10^{-5}$ & 23\,037\,614  &  -0.419\,193    & -0.420\,903(1)   & 541 & 866 \\
$5 \times 10^{-6}$ & 54\,367\,230  &  -0.419\,958    & -0.420\,924(0)   & 1625 & 1346 \\
\bf 0 \tnote{a} &&&  \bf\fontsize{2.7mm}{0}\selectfont -0.420\,934(5) \\
0 \tnote{b} & & &                                      -0.421\,07  \\
\noalign{\vskip 3mm}
& & \multicolumn{2}{c}{Excit. space (24e,30o)} \\
& & \multicolumn{2}{c}{CAS(12e,12o) core} \\
& & \multicolumn{2}{c}{CAS d2h orbitals } \\
$5 \times 10^{-4}$ & 123\,144      & -0.387\,661 &     -0.420\,094(10)  & 1 & 37 \\
$2 \times 10^{-4}$ & 480\,138      & -0.402\,856 &     -0.420\,541(7)   & 5 & 64 \\
$1 \times 10^{-4}$ & 1\,276\,421   & -0.409\,702 &     -0.420\,927(6)   & 14 & 100 \\
$5 \times 10^{-5}$ & 3\,306\,031   & -0.414\,146 &     -0.421\,151(4)   & 45 & 176 \\
$2 \times 10^{-5}$ & 11\,254\,965  & -0.417\,662 &     -0.421\,308(1)   & 243 & 537 \\
$1 \times 10^{-5}$ & 27\,694\,681  & -0.419\,185 &     -0.421\,355(1)   & 698 & 1254 \\
$5 \times 10^{-6}$ & 66\,679\,956  & -0.420\,114 &     -0.421\,375(1)   & 2288 & 1848 \\
\bf 0 \tnote{a} &&&  \bf\fontsize{2.7mm}{0}\selectfont -0.421\,385(5) \\
0 \tnote{b} & & &                                      -0.421\,52  \\
\noalign{\vskip 2mm}
iCIPT2 \cite{ZhaLiuHof-ARX-19} & CAS core & -0.416\,130& -0.421\,470(16) \\
ASCI \cite{TubLeeTakHeaWha-JCP-16} & unknown core & -0.403\,88 & -0.420\,3 \\
ASCI \cite{TubFreLevHaiHeaWha-ARX-18} & unknown core &             & -0.420\,517 \\
DMRG \cite{OliHuNakShaYanCha-JCP-15} & HF core & -0.420\,78 & -0.420\,948(34) \\
FCIQMC \cite{BooSmaAla-MP-14} & HF core & &                   -0.421\,2(3) \\
DMRG \cite{ShaCha-JCP-12}   & unknown core & -0.420\,82 &     -0.421\,00 \\
DMRG \cite{KurYan-JCP-09}   & HF core & -0.420\,525 &         -0.421\,156 \\
CCSDTQ \cite{OliHuNakShaYanCha-JCP-15} & HF core & &          -0.406\,696 \\
\noalign{\vskip 3mm}
& & \multicolumn{2}{c}{Excit. space (48e,42o)} \\
& & \multicolumn{2}{c}{CAS d2h orbitals} \\
$5 \times 10^{-4}$ & 190\,937      & -0.405\,001 &     -0.442\,899(10)  & 3 & 278 \\
$2 \times 10^{-4}$ & 787\,919      & -0.422\,163 &     -0.443\,463(7)   & 13 & 293 \\
$1 \times 10^{-4}$ & 2\,237\,828   & -0.430\,093 &     -0.443\,908(4)   & 40 & 569 \\
$5 \times 10^{-5}$ & 6\,171\,642   & -0.435\,388 &     -0.444\,229(4)   & 145 & 1091 \\
$2 \times 10^{-5}$ & 22\,484\,929  & -0.439\,657 &     -0.444\,450(1)   & 846 & 2177 \\
$1 \times 10^{-5}$ & 58\,390\,489  & -0.441\,566 &     -0.444\,525(1)   & 2671 & 2631 \\
$5 \times 10^{-6}$ & 148\,589\,206 & -0.442\,773 &     -0.444\,560(1)   & 10981 & 2994 \\
\bf 0 \tnote{a} &&&  \bf\fontsize{2.7mm}{0}\selectfont -0.444\,586(10)  \\
0 \tnote{b} & & &                                      -0.444\,75  \\
iCIPT2 \cite{ZhaLiuHof-ARX-19} & & -0.435\,048 &     -0.444\,740(41) \\
ASCI \cite{TubLeeTakHeaWha-JCP-16} & & &               -0.443\,25 \\
DMRG \cite{OliHuNakShaYanCha-JCP-15} & & -0.443\,334 &             -0.444\,78(32) \\
CCSDTQ \cite{OliHuNakShaYanCha-JCP-15} & & &             -0.430\,244 \\
\hline
\end{tabular}
\begin{tablenotes}\footnotesize
\item [a] Weighted quartic fit
\item [b] Linear fit to 4 points with $\epsilon_1$ ranging from $2 \times 10^{-5}$ to $2 \times 10^{-4}$ Ha.
\end{tablenotes}
\end{threeparttable}
\end{table}

\newpage

We end by noting that the energy of course
depends on the choice of which orbitals are frozen (HF orbitals or CAS natural orbs) because this changes the active space in which the calculation
is done.  This makes a difference of 0.45 mHa to the converged energy as can be seen in Fig. 1 and Table 1.
On the other hand, once the core orbitals are selected, the choice of which
orbitals to use in the active space (d2h or Lz symmetry, and, natural orbitals or optimized orbitals), and
whether we use the second selection criterion or not, affects only the rate
of convergence; the final converged energy is unchanged to within the stated accuracy of 0.005, 0.005 and 0.01 mHa.

\clearpage

\bibliographystyle{jchemphys}
%\bibliography{all}